\documentclass[twocolumn,twoside,slac]{revtex4}
\usepackage{graphicx}
\usepackage{fancyhdr}
\newcommand{\svc}  {{service} }
\newcommand{\svcs} {{services} }

\begin{document}

\title{The new BaBar Data Reconstruction Control System}


\author{A. Ceseracciu, M.Piemontese}
\affiliation{SLAC, Stanford, CA 94025, USA / INFN Padova, I-35131 Padova, Italy}
\author{F. Safai Tehrani}
\affiliation{INFN Roma, I-00185 Roma, Italy}
\author{P. Elmer}
\affiliation{Princeton University, Princeton, NJ 08544, USA}
\author{D. Johnson}
\affiliation{University of Colorado, Boulder, CO 80309, USA}
\author{T. M. Pulliam}
\affiliation{Ohio State University, Columbus, OH 43210, USA}
\author{(for the BaBar Computing Group)}

\begin{abstract}
The BaBar experiment is characterized by extremely high luminosity, a complex detector, and a huge data volume, with increasing requirements each year. To fulfill these requirements a new control system has been designed and developed for the offline data reconstruction system.\\
The new control system described in this paper provides the performance and flexibility needed to manage a large number of small computing farms, and takes full benefit of OO design. The infrastructure is well isolated from the processing layer, it is generic and flexible, based on a light framework providing message passing and cooperative multitasking. The system is actively distributed, enforces the separation between different processing tiers by using different naming domains, and glues them together by dedicated brokers. It provides a powerful Finite State Machine framework to describe custom processing models in a simple regular language.\\
This paper describes this new control system, currently in use at SLAC and Padova on $\sim$450 CPUs organized in 12 farms.\\
%SLAC-PUB-9873
\end{abstract}

\maketitle

\thispagestyle{fancy}


\section{The BaBar Event Reconstruction and the need for a  Control System}

\subsection {BaBar Prompt Reconstruction}
The Prompt Reconstruction (PR) system is part of the software for the {\it BaBar} experiment. Its tasks are to generate calibrations, and to process the incoming data from the detector performing a complete reconstruction of physical events so that the results will be ready for running physics analysis on them. These two tasks are done in two separate {\it passes}: the Prompt Calibration (PC) pass, and then the Event Reconstruction (ER) pass. \\	
The data obtained from the detector Data Acquisition System (DAQ), are stored in files using a format known as ``extended tagged container'' (XTC file). Those files are then stored in the mass storage system. There is a 1-to-1 correspondence between data runs and XTC files.\\
The Prompt Calibration pass calculates and  stores the new calibrations for each run into the calibration database, built on Objectivity/DB technology \cite{objy}. These calibrations are then used by the Event Reconstruction pass of the run which produces fully reconstructed events and writes them to the event store database. The current production event store is built on Objectivity/DB.\\
The Prompt Calibration and Event Reconstruction passes are processed by farms of typically 32 and 64 CPUs respectively.\\ 
The farm organization is as follows: a dedicated server machine runs for each run to be processed an instance of a tool known as the Logging Manager (LM, see \cite{lm}). It distributes the events read from an XTC file to the actual reconstruction processes running on the farm machines (nodes). The Prompt Calibration farm must process runs in sequential order, as the calibration uses information from the previous runs, while the Event Reconstruction farm is only constrained to process runs that have been calibrated.\\
A detailed discussion about the BaBar Prompt Reconstruction model is given in \cite{prpaper}.

\section{The first Control System}
The first Prompt Reconstruction Control System and its development process were presented in detail in \cite{franzpaper}. We'll add here only some considerations.
\subsection{Automation}
The Reconstruction software provides all the tools necessary to process a run in a computing farm. The amount of sheer manual work needed to do that by directly using the tools makes it impractical to process the huge data rate produced by BaBar.\\
Hence, the basic purpose of the Control System was to automate the manual procedures that would otherwise be required, in order to obtain a system able to process long sequences of data runs without continuous input from human operators.\\
The system was then realized using the most popular tools available on Unix platform: perl and shell scripting.

\subsection{A necessity driven development process}
As the Reconstruction software moved from development stages to production, the relevance of computing and system issues grew larger. The Reconstruction code stayed mainly concerned with physics, and the responsibility of handling everything else was gradually accounted to the Control System.\\
This necessity driven development process caused the control system scripts to grow in complexity. Quick programming approaches, typical of system level scripting, raised manageability issues; local improvements easily caused global disruption; the behavior of the system became difficult to understand and interact with.\\
Even though the system was able to properly satisfy its functionality requirements, these flexibility concerns motivated the design and implementation of a different system.

\section{The new Control System}
The old system taught a rich series of lessons to its developers. Unexpected problems and requirements cropped up throughout its lifetime. This yielded the following set of requirements for the control system:

\begin{itemize}
\item flexible: able to execute a variety of processing models, distributing processes and tasks on different machines as decided by the user;
\item reliable: complete and efficient error and exception handling, which makes it simpler to add/modify parts;
\item extensible: new modules and features can be easily added to the processing models;
\item simple: both for the end user and for the developer, interfaces are kept as simple as possible;
\end{itemize}

To meet these requirements, the Control System is  a highly distributed system avoiding, as much as possible, single points of failure which caused problems in the former control system. Relevant services are duplicated as needed (statically and dynamically) to enhance robustness and performance. The Control System can be described as a dynamic network of distributed services interacting transparently to perform all the data processing related tasks. These tasks can be summarized as follows:

\begin{itemize}
\item Data Flow
\item Data Processing
\item Distributed Networking
\item Finite State Machines
\item System Configuration
\item System Monitoring and Alarms
\item Task Automation
\item User Interface
\end{itemize}

The entity providing cohesion to the system is the Lightweight Processing Framework (LPF). The framework itself only provides essential services (hence the attribute ``Lightweight''): cooperative multitasking, message passing and dynamic module management. Any other core task is delegated to the core framework modules.\\
Any machine used by the Control System runs at least one LPF, and all components of the Control System are coded inside framework modules.
%% NON QUI
%The effort originally headed by Francesco Safai Tehrani started in  December 2001. The deadline for production was the start of data taking, at the beginning of November 2002,  after the detector upgrade and maintanance shutdown.

%
%\subsection{Evolution of requirements}
%computing model, slac/pd, PC/ER...
%boh indeciso se mettere qualche cosa. 

\subsection{Engineering for flexibility: Services}
\label{subsec-Services}
A \svc is a high level design concept meant to separate functionality from actual implementation. The end-user will see the system in terms of the \svcs it provides, without the need for an in depth knowledge of the internal structure.\\
The key concept of services is topological transparency: any service meant to be visible globally in the system can be accessed from any LPF pretending it's local. A transparent proxy is automatically instantiated and handles the delivery to the remote end, as well as the routing of the answers to the message.\\
Also, \svcs can be hierarchically layered in order to create higher level \svcs which are potentially more descriptive of the problem domain and thus able to hide from the user some details of the programming interface.

\section{\label{LPF}Framework and agents}
\subsection{Introducing the LPF}
The LPF is the base element of this distributed architecture. Each node participating in the farm runs at least one LPF agent,  thus allowing for remote activation of services from one or more remote LPFs. This also allows dynamic restructuring of the Control System, by easily ``moving'' \svcs around in a completely transparent fashion.\\
We can see the set of all the LPFs running in the system as a Distributed Service Manager. The most important characteristics of the LPF are: 

\begin{itemize}
\item it is a general purpose framework;
\item it is a \svc controller/organizer;
\item it can easily be configured;
\item it is ``lightweight'' (that is: its features are kept at minimum in order to maximize lightness and generality).
\end{itemize}

The framework has the  ability to ``load'' \svcs and objects through well designed interfaces.  The implementation of the LPF was  the first step toward the implementation of the Control System.\\
Framework functionality is provided by a set of core modules. Dedicated modules take care of:
\begin{itemize}
\item message passing front-end %(see \ref{imc})
\item dynamic module allocation %(see \ref{ma})
\item TCP/IP inter-LPF communication %(see \ref{mpx})
\item Transparent proxy system
\item Fault-tolerant naming service %(see \ref{ns})
\item Active handling for external processes %(see \ref{sysh})
\item Internal event scheduling, cron like
\item Local and global log message handling
\item Alarm handling
\end{itemize}

\subsection{Messages}
Any inter-module communication is mediated by messages, i.e. instances of the \verb OprMessage \ class. Messages are then transparently mapped by the LPF core modules to remote method invocations. Message passing is strictly asynchronous. Routing of messages to different LPFs is handled implicitly via transparent proxies.

\subsection{The Structure of the LPF} 
The LPF is a transparent facility: the operator never needs to interact with it directly.\\
The uniformity of interactions between the framework and the object/\svc modules is guaranteed by the requirement that all the Framework-instantiable objects implement a common interface.\\ %?put the class name here?
The LPF is an  event based server with a simulated multitasking structure, also known as cooperative multitasking. In this processing model there is a single flow of execution in which many parts (modules) participate. Each module must be written in such a way that it performs a single set of operations, in a limited amount of time, and then returns control to the caller.\\
After performing the initialization, the Framework reaches the ``steady state'', that is the main\_event\_loop. During this phase it cyclically transfers control to each one of the modules, thus simulating a simultaneous flow of execution of various applications.\\
%\begin{verbatim}
%        main_event_loop <--
%                \          |
%                 ModuleA   |
%                /          |
%        main_event_loop    |
%                \          |
%                 .....     |
%                /          |
%        main_event_loop    | 
%                \          |
%                 ModuleZ   | 
%                /          |
%        main_event_loop -->
%\end{verbatim}
Since there is a centralized control system, inter-object communication becomes completely transparent without the need for any additional complex inter process communication structure. Modules interact via messages. In particular, each module generates  a list of messages to be distributed to other modules as a return value.\\
%\begin{verbatim}
%        main_event_loop <--
%                \          |
%                 ModuleA   |
%                /          |
%        main_event_loop    |
%                \          |
%                 .....     |
%                /          |
%        main_event_loop    | 
%                \          |
%                 ModuleZ   |
%                /          |
%        main_event_loop    |
%                \          |
%           MsgDistribution | <--- takes the FrameworkMsgQueue and
%                /          |      distributes it to the Modules.
%        main_event_loop -->
%\end{verbatim}

The modules in the Framework can be logically divided into two separate groups:
\begin{itemize}
\item Active Modules: these need to be given control during every iteration.
\item Passive Modules: these are given control only when they receive a message.
\end{itemize}

The module interface provides simple methods to allow the Framework to transparently interact with the modules:

\begin{itemize}
\item Do: receives a message and executes the associated action  returning the answer as one or more messages;
\item Kill: performs the set of operation needed to remove cleanly a module from the system;
\item Init: invoked after instantiation, allows modules to execute a custom startup sequence, eventually based on the configuration.
\item Run: performs a set of operations to be executed during every iteration for Active Modules. Passive Modules don't define this method.
\end{itemize}

The main\_event\_loop of the LPF looks like this:
\begin{itemize}
\item begin\_event\_loop:
\item invoke the run method of each active module and ...
\item ...load the messages returned in the LpfMsgQueue; 
\item invoke the run method of the InterModuleCommunication passing the LpfMsgQueue as an argument;
\item the InterModuleCommunication dispatches the messages to the receivers and...
\item ...collects the answer messages that they generate and puts them back on the LpfMsgQueue;
\item end\_event\_loop;
\end{itemize}

\subsection{Core Modules}
Some interesting core modules are presented in this section.
\subsubsection{\label{mpx}MpxServer}
The \verb MpxServer \ service provides the LPF server network interface. It is realized as a multiplexed server: all the clients connect to the main loop and their messages are collected by cyclic polling. All the queries from the clients are streamlined to the InterModuleCommunication module to be distributed to the relevant modules.\\
The \verb MpxServer \ service also automatically becomes the intermediate destination for all the messages sent back to the clients as answers to the client queries, that is, it acts as a virtual Proxy for all the incoming client connections.

\subsubsection{\label{imc}InterModuleCommunication}
The \verb InterModuleCommunication \ module  takes care of dispatching all the messages passed to an LPF to the correct destination modules. It provides various logging levels for debugging purposes.\\
It also provides the ability to choose what to do with messages for which a receiver is not available, discarding them quietly or with an alarm, and the ability to instantiate modules on request in case they are not available in the system, so that those messages can then be dispatched. The \verb ModuleActivator \ module  is used to perform the actual instantiation and passing of the configuration to the module.

\subsubsection{\label{ma}ModuleActivator}
Dynamic loading and instantiation of modules is the duty of the \verb ModuleActivator . Beyond the actual loading and configuration of local modules, the main duty of the \verb ModuleActivator \ is the transparent handling of global services.\\
When activation of a global service is requested, a lookup is made to the naming Service; if the service is already running in some other LPF in the same naming domain, a transparent proxy to it is activated in place.

\subsubsection{\label{ns}NamingService}
The Naming Service is a replicated server with a broker layer. Replicated servers can be configured to run on different machines, and contain the same information. They index service names, and store their current location. The location is the pair (hostname, port) that identifies a running LPF.\\
The broker service is designed to be restartable at any time without disrupting the Naming Service. It serves as an access point to the service, and takes care of data replication and load balancing between servers. It does not store any dynamic data, except current transactions.

\subsubsection{\label{sysh}OprSysHandler}
The \verb OprSysHandler \ module handles the running of
external processes, like the actual reconstruction tool (Elf), or the LoggingManager.  It can be used to execute any external process without worrying that the process will block or crash the LPF.\\
The actual running of a particular process is handled by another object,
namely \verb OprProcess . \verb OprSysHandler \ is the interface to
\verb OprProcess , therefore hiding all the gory details of running and
managing processes from the client.\\
A process is started when a client sends a \verb Start \ message to the
\verb OprSysHandler . The client is then returned a message containing a
reference to an hash, which the client can use to examine the status of
the process. The client can also send additional
messages to the \verb OprSysHandler \ to  determine the status of the process,
retrieve any output the process generated, and to
signal (KILL, SUSPEND, CONTINUE) the process, if needed.

\subsection{The Agents deployment}
\begin{figure*}[t]
\centering
\includegraphics[width=140mm]{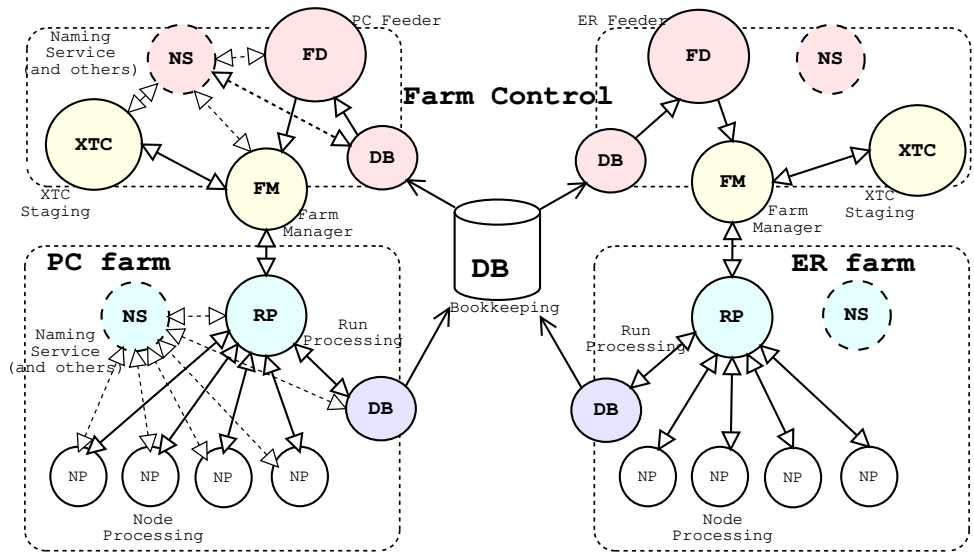}
\caption{The actual deployment of the Control System agents, and the partitioning between multiple farms and farm subsystems.} 
\label{farmcontrol}
\end{figure*}

  Each LPF agent of the new Control System can be configured to dynamically load a number of \svcs. Figure~\ref{farmcontrol} shows the architecture of the Control System for both the Prompt Calibration (PC) and the Event Reconstruction (ER) pass farms, with emphasis on the communication flow.\\
The main structure of both passes is very similar; the main driver of the Control System is the Farm Manager (FM) \svc that acts as a broker between two separate levels of service: the first level includes all  \svcs  (upper part of figure~\ref{farmcontrol}) needed for the staging (XTC) to disk of the runs and the \svcs that take care of the scheduling of the runs; the second level (bottom part of figure~\ref{farmcontrol}) includes all the services that take care of the event reconstruction (ER pass) or calibration (PC pass). The Farm Manager  serves also  as the main access point for the user to the system. The two layers have separate  services' namespaces, each one of them having a Naming Service (NS).\\
The reconstruction part of the single run is controlled by a Run Processing (RP) Finite State Machine \svc (see section~\ref{fsm} for the Finite State Machine abstraction and \svc implementation) that starts the Logging Manager \cite{lm} on the farm server and takes care of the bookkeeping; it also starts and dynamically configures  a Node Processing (NP) Finite State Machine \svc to handle each Elf reconstruction process. Typically each CPU runs an Elf.\\
Each of the Node Processing FSM (see figure~\ref{NodeProcessingFSM} for the description of the states of the NP) starts the Elf and waits until it has finished, then it checks for outstanding locks on the event store database and gathers information from the output of the Elf. Each layer of the Control System has an interface \svc to the bookkeeping database. 

\subsection{The unified configuration and activation system}
\label{subsec-config}
\subsubsection{Formal hierarchic organization}
The configuration system is designed to define the whole configuration of a large and complex computing system in a unique structure, contained in a small number of files. This centralized approach makes life easier for the user, but also is a potential performance and reliability problem, since it is a centralized service in a distributed environment.\\
A formal hierarchical model allows the maintenance of a global configuration, but grants the flexibility to configure any group of computing nodes with the desired granularity. An important advantage of a formal hierarchy is that it gives the computing environment a controlled and well defined structure. A strong structure is the only way to harness the complexity of a large and heterogeneous system, as PR. A clear example is how the activation system benefits from the structure specified at configuration level, as described below.

\subsubsection{Initialization system}
Initialization must be completed before Activation.\\
The Control System initialization procedure has been reduced to just one operation: running on every node a special purpose LPF, called ``BareLPF''.\\
A BareLPF is an LPF which only runs some core modules. Its configuration is extremely simple and entirely hard-coded in its special startup file. The BareLPF listens on a fixed TCP port, which should be reserved for this service. The BareLPF does not need any other external information or resources to run.\\
The one and only duty of a BareLPF is to spawn new LPFs. A new LPF is spawned upon reception of a specific message.\\
The BareLPF may be started at boot time or remotely by means of ssh.

%this may be omitted
\subsubsection{\label{activation}Distributed Activation Protocol}
Thanks to the hierarchical model of computation, we will just define the activation protocol for one generic case: a Configuration Manager (A) which activates a lower-level Configuration Manager (B). The activation of the whole computing environment follows by induction.\\
From a local point of view, configuration is performed after activation. On the whole system scale, though, activation and configuration are performed at the same time, because the hierarchy information, needed for the activation, is specified in the configuration.\\  
The configuration sequence of an LPF includes the activation of all its children in the hierarchy tree, and cannot return until all its children report success or failure, or after its timeout expires.\\
Depending on what is specified in the configuration, failure to activate a child LPF can lead to the failure of the parent's configuration sequence, or can be ignored, or recovered.\\
When the configuration sequence of the Configuration Master LPF (the top level Configuration Manager) ends successfully, the whole computing system is guaranteed to be configured and ready for operation.
 
\subsubsection{Configuration parsing and propagation}
The configuration file is valid globally. New spawned LPFs would be able to read the global configuration file, parse it, and find their own subtree.\\
This approach would require the configuration file to be distributed to, and parsed by, every node. This can be clearly a bottleneck from the network point of view; the main disadvantage though is that the distribution of the configuration files would require some software component different from the LPF, like nfs, ssh, ftp, i.e. one more potential problem.\\
In the activation and configuration protocol design the configuration files are parsed only once, by the Configuration Master. The parsed configuration structure is passed to the child LPFs by the AnswerRequestConfig method. This guarantees that all the nodes configuration is coherent, i.e. built from the same files.\\

\subsubsection{Local configuration}
The \verb OprConfigurationService \ recognizes the configuration keys relevant for its LPF, and makes them available to the LPF itself and its modules.\\
The information that is valid for the whole LPF, i.e. the keys specified in the LpfConfig section of the configuration file, is copied to a dedicated structure inside the local LPF context.\\
The configuration supplied for a module is directly copied to a dedicated structure in the data space of the module itself, so that the module needs not to make explicit access to the global configuration structure.

\subsubsection{System deactivation}
The activation service is also responsible for the deactivation. The deactivation subsystem is also known as the Reaper.\\
The deactivation information is maintained on each node by its bareLPF, specifically, by an instance of the \verb OprLocalLpfMap \ module loaded on the bareLPF.\\
When the OprLocalLpfMap sitting on a bareLPF receives the LocalLpfReaper command, it checks if it carries the name of the Service being reaped. If it does not, the bareLPF sends to each of the LPFs active on its host a StopLocalLpf message; otherwise, only LPFs belonging to the specified service are targeted.\\
An LPF upon reception of the StopLocalLpf runs a self-destruction sequence: it stops and unloads its modules and finally exits the main LPF loop.
The bareLPF remains active and ready to undergo a new activation of the system.\\
The deactivation sequence can be run from any LPF in which the full configuration has been loaded; this LPF becomes the Deactivation Master. It is possible to deactivate the whole system at once, i.e. to stop all LPFs running on some node defined in the configuration, even if not bound to a configured service; another option is to deactivate a single System or Service defined in the configuration.

\subsection{\label{fsm}The Finite State Machine framework}

A Finite State Machine (FSM) is a computation model consisting of a discrete set of states, a start state, an input alphabet, and a transition function which maps input symbols and current states to a next state. Computation begins in the start state with an input string.  It changes to new states depending on the transition function. 

\begin{figure*}[t]
\centering
\setlength{\unitlength}{2763sp}%
\begingroup\makeatletter\ifx\SetFigFont\undefined%
\gdef\SetFigFont#1#2#3#4#5{%
  \reset@font\fontsize{#1}{#2pt}%
  \fontfamily{#3}\fontseries{#4}\fontshape{#5}%
  \selectfont}%
\fi\endgroup%
\begin{picture}(9300,4341)(976,-4390)
{\thinlines
\put(4201,-3511){\circle*{150}}
}%
{\put(4201,-3511){\circle{336}}
}%
\put(3451,-2236){\makebox(0,0)[b]{\smash{\SetFigFont{8}{9.6}{\rmdefault}{\mddefault}{\updefault}{generate}%
}}}
\put(3451,-2461){\makebox(0,0)[b]{\smash{\SetFigFont{8}{9.6}{\rmdefault}{\mddefault}{\updefault}{alarm}%
}}}
\put(3451,-661){\makebox(0,0)[b]{\smash{\SetFigFont{8}{9.6}{\rmdefault}{\mddefault}{\updefault}{ElfOprApp}%
}}}
\put(3451,-436){\makebox(0,0)[b]{\smash{\SetFigFont{8}{9.6}{\rmdefault}{\mddefault}{\updefault}{start}%
}}}
\put(6151,-436){\makebox(0,0)[b]{\smash{\SetFigFont{8}{9.6}{\rmdefault}{\mddefault}{\updefault}{}%
}}}
\put(6151,-661){\makebox(0,0)[b]{\smash{\SetFigFont{8}{9.6}{\rmdefault}{\mddefault}{\updefault}{EventProcessing}%
}}}
\put(8851,-436){\makebox(0,0)[b]{\smash{\SetFigFont{8}{9.6}{\rmdefault}{\mddefault}{\updefault}{collect}%
}}}
\put(8851,-661){\makebox(0,0)[b]{\smash{\SetFigFont{8}{9.6}{\rmdefault}{\mddefault}{\updefault}{Log statistics}%
}}}
\put(6151,-3436){\makebox(0,0)[b]{\smash{\SetFigFont{8}{9.6}{\rmdefault}{\mddefault}{\updefault}{clearup}%
}}}
\put(6151,-3661){\makebox(0,0)[b]{\smash{\SetFigFont{8}{9.6}{\rmdefault}{\mddefault}{\updefault}{locks}%
}}}
{\put(6864,-3961){\oval(526,456)[bl]}
\put(6864,-3924){\oval(530,530)[br]}
\put(6901,-3924){\oval(456,526)[tr]}
\put(6901,-3661){\vector(-1, 0){0}}
}%
\put(7576,-4336){\makebox(0,0)[b]{\smash{\SetFigFont{8}{9.6}{\rmdefault}{\mddefault}{\updefault}{Stay}%
}}}
{\put(4164,-961){\oval(526,456)[bl]}
\put(4164,-924){\oval(530,530)[br]}
\put(4201,-924){\oval(456,526)[tr]}
\put(4201,-661){\vector(-1, 0){0}}
}%
\put(4876,-1336){\makebox(0,0)[b]{\smash{\SetFigFont{8}{9.6}{\rmdefault}{\mddefault}{\updefault}{Stay}%
}}}
{\put(6864,-961){\oval(526,456)[bl]}
\put(6864,-924){\oval(530,530)[br]}
\put(6901,-924){\oval(456,526)[tr]}
\put(6901,-661){\vector(-1, 0){0}}
}%
{\put(9564,-961){\oval(526,456)[bl]}
\put(9564,-924){\oval(530,530)[br]}
\put(9601,-924){\oval(456,526)[tr]}
\put(9601,-661){\vector(-1, 0){0}}
}%
{\put(1201,-511){\circle{150}}
}%
{\put(2806,-856){\oval(210,210)[bl]}
\put(2806,-166){\oval(210,210)[tl]}
\put(4096,-856){\oval(210,210)[br]}
\put(4096,-166){\oval(210,210)[tr]}
\put(2806,-961){\line( 1, 0){1290}}
\put(2806,-61){\line( 1, 0){1290}}
\put(2701,-856){\line( 0, 1){690}}
\put(4201,-856){\line( 0, 1){690}}
}%
{\put(2401,-511){\makebox(2.3810,16.6667){\SetFigFont{5}{6}{\rmdefault}{\mddefault}{\updefault}.}}
}%
{\put(2806,-2656){\oval(210,210)[bl]}
\put(2806,-1966){\oval(210,210)[tl]}
\put(4096,-2656){\oval(210,210)[br]}
\put(4096,-1966){\oval(210,210)[tr]}
\put(2806,-2761){\line( 1, 0){1290}}
\put(2806,-1861){\line( 1, 0){1290}}
\put(2701,-2656){\line( 0, 1){690}}
\put(4201,-2656){\line( 0, 1){690}}
}%
{\put(5506,-856){\oval(210,210)[bl]}
\put(5506,-166){\oval(210,210)[tl]}
\put(6796,-856){\oval(210,210)[br]}
\put(6796,-166){\oval(210,210)[tr]}
\put(5506,-961){\line( 1, 0){1290}}
\put(5506,-61){\line( 1, 0){1290}}
\put(5401,-856){\line( 0, 1){690}}
\put(6901,-856){\line( 0, 1){690}}
}%
{\put(1276,-511){\vector( 1, 0){1425}}
}%
{\put(3301,-961){\vector( 0,-1){900}}
}%
{\put(8206,-856){\oval(210,210)[bl]}
\put(8206,-166){\oval(210,210)[tl]}
\put(9496,-856){\oval(210,210)[br]}
\put(9496,-166){\oval(210,210)[tr]}
\put(8206,-961){\line( 1, 0){1290}}
\put(8206,-61){\line( 1, 0){1290}}
\put(8101,-856){\line( 0, 1){690}}
\put(9601,-856){\line( 0, 1){690}}
}%
{\put(4201,-511){\vector( 1, 0){1200}}
}%
{\put(6901,-511){\vector( 1, 0){1200}}
}%
{\put(5401,-3511){\vector(-1, 0){1050}}
}%
{\put(3301,-2761){\line( 0,-1){750}}
\put(3301,-3511){\vector( 1, 0){750}}
}%
{\put(6001,-961){\line( 0,-1){1125}}
\put(6001,-2086){\vector(-1, 0){1800}}
}%
{\put(9001,-961){\line( 0,-1){2550}}
\put(9001,-3511){\vector(-1, 0){2100}}
}%
{\put(5506,-3856){\oval(210,210)[bl]}
\put(5506,-3166){\oval(210,210)[tl]}
\put(6796,-3856){\oval(210,210)[br]}
\put(6796,-3166){\oval(210,210)[tr]}
\put(5506,-3961){\line( 1, 0){1290}}
\put(5506,-3061){\line( 1, 0){1290}}
\put(5401,-3856){\line( 0, 1){690}}
\put(6901,-3856){\line( 0, 1){690}}
}%
{\put(6001,-3061){\line( 0, 1){450}}
\put(6001,-2611){\vector(-1, 0){1800}}
}%
{\put(8401,-961){\line( 0,-1){1350}}
\put(8401,-2311){\vector(-1, 0){4200}}
}%
\put(976,-286){\makebox(0,0)[b]{\smash{\SetFigFont{8}{9.6}{\rmdefault}{\mddefault}{\updefault}{StartProcessing}%
}}}
\put(4876,-2011){\makebox(0,0)[b]{\smash{\SetFigFont{8}{9.6}{\rmdefault}{\mddefault}{\updefault}{ERROR}%
}}}
\put(8026,-3436){\makebox(0,0)[b]{\smash{\SetFigFont{8}{9.6}{\rmdefault}{\mddefault}{\updefault}{OK}%
}}}
\put(3676,-1561){\makebox(0,0)[b]{\smash{\SetFigFont{8}{9.6}{\rmdefault}{\mddefault}{\updefault}{ERROR}%
}}}
\put(4801,-436){\makebox(0,0)[b]{\smash{\SetFigFont{8}{9.6}{\rmdefault}{\mddefault}{\updefault}{OK}%
}}}
\put(7501,-436){\makebox(0,0)[b]{\smash{\SetFigFont{8}{9.6}{\rmdefault}{\mddefault}{\updefault}{OK}%
}}}
\put(6676,-2236){\makebox(0,0)[b]{\smash{\SetFigFont{8}{9.6}{\rmdefault}{\mddefault}{\updefault}{ERROR}%
}}}
\put(4876,-2836){\makebox(0,0)[b]{\smash{\SetFigFont{8}{9.6}{\rmdefault}{\mddefault}{\updefault}{ERROR}%
}}}
\put(3001,-3211){\makebox(0,0)[b]{\smash{\SetFigFont{8}{9.6}{\rmdefault}{\mddefault}{\updefault}{END}%
}}}
\put(4876,-3436){\makebox(0,0)[b]{\smash{\SetFigFont{8}{9.6}{\rmdefault}{\mddefault}{\updefault}{OK}%
}}}
\put(7576,-1336){\makebox(0,0)[b]{\smash{\SetFigFont{8}{9.6}{\rmdefault}{\mddefault}{\updefault}{Stay}%
}}}
\put(10276,-1336){\makebox(0,0)[b]{\smash{\SetFigFont{8}{9.6}{\rmdefault}{\mddefault}{\updefault}{Stay}%
}}}
\end{picture}
\caption{A simplified view of the NodeProcessingFSM}
\label{NodeProcessingFSM}
\end{figure*}
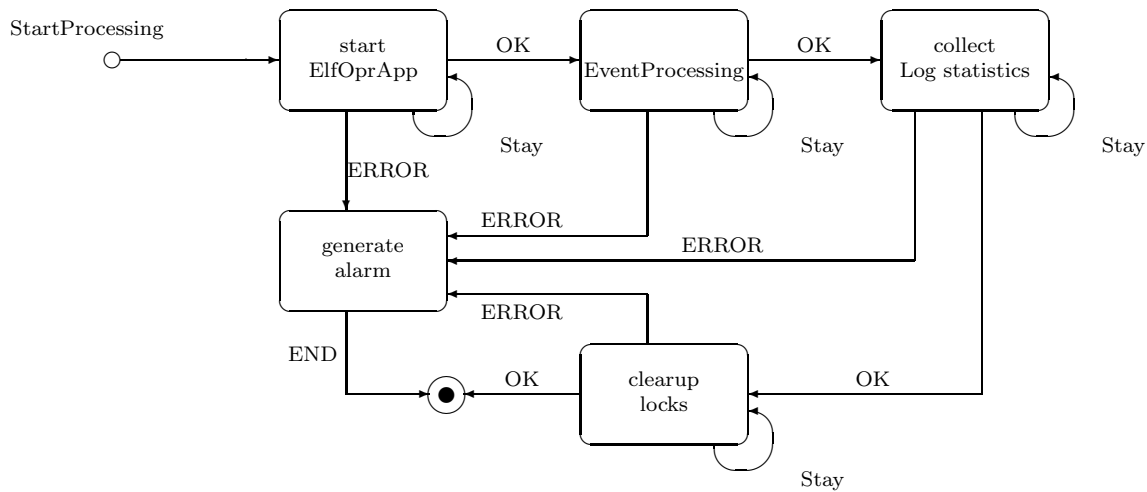

The processing structure of PR can be easily described in terms of states and transitions between the states determined by well defined conditions. This offers a very flexible model to describe a processing system and realize it. Figure \ref{NodeProcessingFSM} shows succession of the operations (states) and transitions between them of the NodeProcessingFSM.\\
The FSM can be considered a specialized, heavyweight framework dedicated to the application programming. All the services and facilities provided by the core system, and almost any application code is encouraged to be coded inside FSM states, are made available through the FSM.

\subsubsection{The FSM Structure}
A generic implementation of an FSM is contained in the \verb OprFSMFwkModule \ class. This class implements the LPF module interface, thus making it possible to plug an FSM in an LPF at run time. FSMs are  specialized  dynamically by loading a particular FSM definition to an instance of the \verb OprFSMFwkModule \ class.

\subsubsection{The FSM Interpreter and the FSM Description Language}
\label{FSM_dl}
The FSM Description Language is a simple description language designed to describe, extend and modify FSMs with the minimum effort. It is stackless, stateless and doesn't have any control structure, only assignments and logical connectors.\\
%%non so se voglio mettere sta roba
The language grammar is very simple:

\begin{itemize}
\item {\tt FSM: [FSM name]}: defines the symbolic name of the FSM;
\item {\tt begin: [state name]}: defines the name of state where the FSM starts;
\item {\tt state: [state name] isA: [class name]}: associates a symbolic name to a class name for the state;
\item {\tt state: [state name] onTransition: [transition name] do: [method name]}: defines a method name to be called on the object that represents the class when a certain transition is generated;
\item {\tt state: [state name] onEntry: [method name]}: defines a method to be invoked on the state object upon entering in the state;
\item {\tt state: [state name] onExit: [method name]}: defines a method to be invoked on the state object upon exiting the state;
\item {\tt state: [state name] timeout: [seconds]}: associates a generic alarm timeout to this state, to spot blocked processing early.

\end{itemize}

\subsection{BaBar Reco Finite State Machines}
%slide 10, long......
%also talk here about the modules not yet presented
The RunProcessing (RP) FSM describes the centralized part of the processing model concerning a given run.\\
For each processed run, a new instance of the RunProcessing FSM is created, configured, and started. This implicates that any memory of the previous runs is erased, hence ensuring reproducibility of the processing, an important consistency requirement. Together with the FSM, all its states are reloaded.\\
Most of the logic of the Control System is coded as states of this FSM, or as modules directly called by proxy states. The complexity of actual RP FSMs is thus considerable, amounting to up to 40 different states.\\
The current status of the RunProcessing FSM is often the best indication of the status of the whole processing farm. It allows the operator to monitor the processing by just observing the transitions between states. A tunable alarm timeout associated with each state is also an effective way of spotting problems that are not caught by specific alarm checks.\\
The RunProcessing FSM is remotely instantiated, configured and started by the FarmManager as soon as all the information needed to start processing a run is available.\\
The first states of the RunProcessing FSM are devoted to collection of information from different sources and consistency checks; a new directory is created to host all the logfiles for the current run. Then, the Logging Manager (\cite{lm}) is started, and a monitor is attached to its log files. The dynamic configuration for the reconstruction processes (Elves) is then produced and written on disk. At this point, the duty of local processing on the nodes is delegated to the NodeProcessing FSMs running on them. They are started and the relevant part of the run-time configuration is passed to them. The RP then just waits for all of them to return. This is the phase where the farms spend most of their time, and where the nodes are actually used for distributed computation.\\
After that, the RunProcessing FSM resumes control and takes care of consistency checking, persistent bookkeeping, and postprocessing, which includes starting the quality assurance procedures. When the final state is reached, the RunProcessing FSM instance returns control to the FarmManager, and is reset before the next run.
%... altro?

\begin{figure*}[t]
\centering
\includegraphics[width=145mm]{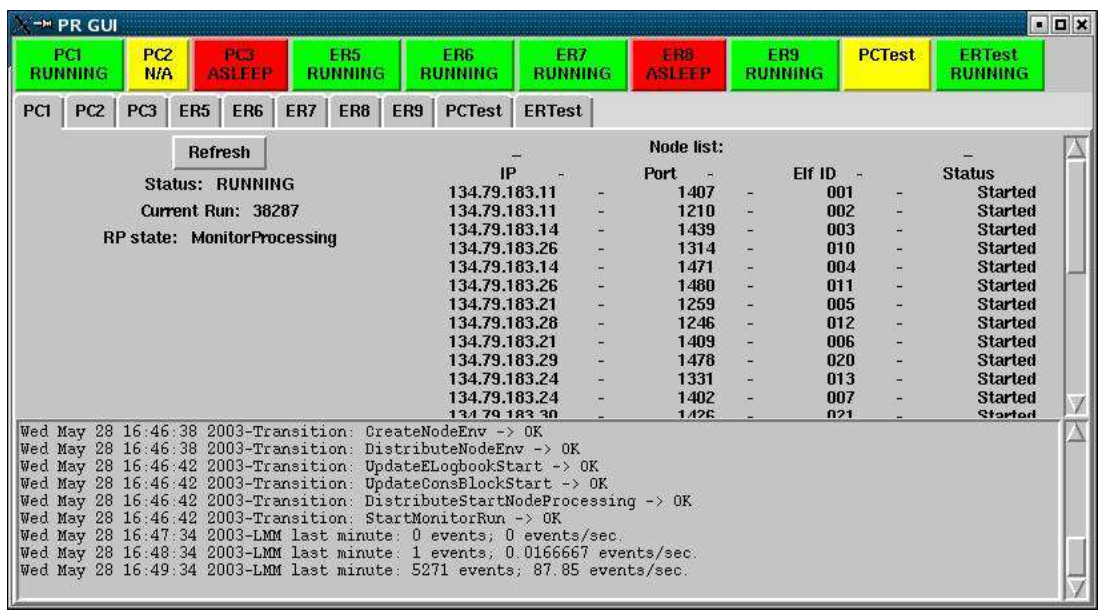}
\caption{The Control System GUI monitoring the processing farms at SLAC}
\label{csgui}
\end{figure*}

\subsection{User Interface}
%cmd-line / shell / gui
The user interface makes no use of the LPF framework and modules. It is nevertheless designed modularly to assist flexibility and maintenance.\\
The user interface and the control system communicate via messages. User commands prepare messages and deliver them to some LPF. Higher level commands use a discovery protocol, which relies on the bareLPF, to locate the destination LPFs.\\
The modular design makes the whole set of commands available by different means; currently provided are a command line interface and an integrated shell. Any command is automatically available from both.\\ 
A further flexibility mechanism is supplied to code different sets of commands, called ``interfaces'', into separate classes. A selection of interfaces can be loaded and changed at run-time in the command executable. Specific interfaces are dedicated to the highest level commands (\verb User ), farm administration (\verb Farm), debugging (\verb Debug ). The ability to send any kind of message to any module makes the command interface a powerful tool for interactive debugging and unit testing.\\ 
 The Control System GUI is a monitoring tool. Its main panel is displayed in figure \ref{csgui}. It allows to inspect at a glance a set of farms, giving immediate visual feedback about the current status of each. More detailed information about a single farm is available at a mouse click, including the current RunProcessing FSM state and the nodes status (when applicable), and a continuously updated log messages history. Multiple instances of the GUI can run at the same time, even on different machines. Any communication between the GUI and the Control System is carried by regular messages. The GUI uses the perl/Tk graphics toolkit
and simulates the LPF main loop thread to implement asynchronous updating of the farms status.

\section{Project Report}
%slide 11: timescale, people, considerations...
The project to design and implement the new Control System started in January 2002, under the technical lead of Francesco Safai Tehrani. The project was then gradually taken over by the two main developers, Antonio Ceseracciu and Martino Piemontese.\\
As the main core parts were completed, the coding of the application part started alongside, in August 2002. A separate package was devoted to this higher layer.\\
The system wasa ready for production in November 2002, in time to be used for the processing of BaBar Run3, started on November 03. There were still many missing pieces at the time, mostly at the user interface level, but the system was able to carry out its main duties.\\
The major remaining components were implemented by Mar 2003, along with a graphical user interface. Maintenance of the system and minor improvements continued afterwards, and were carried out mainly by the original developers.\\
In March 2003 the system was ported to run at the INFN computing facility in Padova. The porting process mainly involved writing a new configuration and a few alternative modules, to properly adapt to the different hardware and software environment.\\
More development effort will take place to adapt the system to properly handle the different requirements of the new BaBar computing model, bearing a major change in the eventstore format. This effort will mainly concern the higher layers of processing code, and will be directed towards the creation of new, alternative states for the FSMs rather than changing the existing code.

\section{Conclusions}

\subsection{Performance}
The Control System is not computation intensive software. This fundamental assumption supports the choice of Object Oriented (OO) programming with a dynamically typed, interpreted language. This combination leads to remarkably slower code than in non-OO fashion, as much as 50\% for \em perl \em, because of the need for the interpreter to resolve method calls at run-time, rather than during bytecode compilation.\\
A more significant performance measure for the Control System is, rather, reliability.\\
The experience of about six months of continuous running shows that the time lost for failure of the Control System itself is a small fraction of the total. More frequently, changes in the operating environment, like new revisions of tools, created conditions that the Control System was not ready to deal with, until properly adapted (``fixed''). This is inherent to the glue-like nature
of the system itself.

\subsection{Software design}
The Control System is an active distributed system. This is a complex programming paradigm, that requires careful design and attention to many issues typical in concurrent programming.\\
This is not unnecessary complexity however. The main task of the Control System, launching and monitoring external executables, simply requires an active monitor. The only passive way to perform this task, even for simple cases like determining when an external command execution ends, is to poll for lock files and eventually parsing the command logfiles for specific messages. This approach is less elegant and reliable: as an example, in a distributed environment it introduces a considerable burden, the dependency from the quasi-coherence (hence in-coherence) born by the distributed file system.\\

Different software engineering techniques were exploited to manage this complexity. The main focus has been on attaining modularity by design.\\
A very light framework (LPF) hosts all of the system, including core parts, like TCP-IP intercommunication and module activation. Hence, anything coded for the Control System has to run inside a framework module. A more structured framework, based on a Finite State Machine abstraction, supports the final processing code. This support makes it convenient to exploit the modular interface even for apparently trivial tasks, thus fostering the developers to design modular code. The main operative advantage of this design is the ease of adapting the system to changing environments.\\
The programming interface of the framework and of most core components has remained stable over time. The maintenance and extension of the system are made much easier by the underlying frameworks and services, and don't cause the code base to bloat with cross-dependencies.


\end{document}